\documentclass[aps,prl,twocolumn,showpacs,floatfix]{revtex4}

\usepackage{graphicx}
\usepackage{amsfonts}
\usepackage[figuresright]{rotating}  
\usepackage{amssymb}
\usepackage{amsmath}
\usepackage{psfrag}
\usepackage{subfigure}
\usepackage{multirow}
\usepackage{tabularx}
\usepackage{textcomp}
\usepackage{units}

\def\nn{\nonumber}

\def\beq{\begin{eqnarray}}
\def\eeq{\end{eqnarray}}
\def\c{\hspace{2pt}}

\renewcommand{\v}[1]{\ensuremath{\mathbf{#1}}} 
 
\newcommand{\uv}[1]{\ensuremath{\mathbf{\hat{#1}}}} 
\let\baraccent=\= 
\renewcommand{\=}[1]{\stackrel{#1}{=}} 

\begin{document}

\title{Ferro-Nematic ground state of the dilute dipolar Fermi gas}
\author{Benjamin M. Fregoso and Eduardo Fradkin}
\affiliation{Department of Physics, University of Illinois, 1110 West Green Street, Urbana, Illinois 61801-3080, USA}
\date{\today}
\begin{abstract}
It is shown that a homogeneous two-component Fermi gas with (long range) dipolar and short-range isotropic interactions has a 
{\em ferro-nematic} phase for suitable values of the dipolar and short-range coupling constants. 
The ferro-nematic phase is characterized by having a non-zero magnetization and 
long range orientational uniaxial order. The Fermi surface of spin up (down) 
component is elongated (compressed) along the direction of the magnetization. 
\end{abstract}
\maketitle 

Cold dipolar Fermi gases have attracted much attention due to the novel anisotropic 
and long-range character of dipole-dipole interactions. Recent studies of many-body effects 
predicted an elongated Fermi surface (FS) in a one-component fully polarized Fermi
gas with dipolar interactions along the polarization direction established by an 
external field\cite{Miyakawa2008,Sogo}. A biaxial state 
with a critical value of the effective coupling constant $\lambda_d=n \mu^2 /\epsilon_F$ 
was proposed in Ref. \cite{Fregoso-2009} ($\mu$ is the dipole moment of the 
fermion, $n$ is the total density and $\epsilon_F$ is the Fermi energy of the 
free Fermi gas at the same density).   It was found\cite{Fregoso-2009} 
that the system will exhibit violations of the Landau theory of the Fermi 
liquid both at quantum criticality and in the biaxial phase. More generally, 
understanding anisotropic non-Fermi liquid phases of cold atomic systems may 
shed light into the quantum liquid crystal phases in strongly correlated 
systems and high $T_c$ superconductors \cite{Kivelson1998,Oganesyan2001,Sun2008}.

The question we want to address here is whether a cold spin-\nicefrac{1}{2} Fermi gas with long range dipolar interactions
can become spontaneously polarized and what is the nature of the broken symmetry state. The theory we present here is a generalization of
the theory of the Stoner (ferromagnetic) transition in metals\cite{Doniach-1974} to take into account 
the effects of the long range and anisotropic dipolar interaction\cite{mahanti-comment}. As we will see below, much as in the theory of Stoner
ferromagnetism, the polarized state can occur only for sufficiently large values of the magnetic dipole moment and/or of the spin-flip scattering rate.
However, unlike what happens in Stoner ferromagnetism, as a result of the structure of the dipolar interactions, 
the resulting polarized state is also spatially anisotropic, a {\em ferro-nematic} state.

The {\em classical} version of this problem has been considered in mixtures 
of ferromagnetic particles with nematic liquid crystals\cite{Brochard1970}, 
and in dipolar colloidal fluids and ferrofluids\cite{seul-1995}. Classical 
dipolar fluids have complex phase diagrams, typically featuring inhomogeneous 
phases with complex spatial structures. Much less is known about their 
quantum counterparts.  In the case of simple quantum fluids, such as  
${}^3$He, the dipolar interaction plays a small role compared to the 
short-range exchange interaction\cite{Fomin1978}. In the context of 
ultracold gases, a number of atomic and molecular systems with strong 
dipolar interactions, such as Dy, have been the focus of recent 
experiments 
(see Ref.\cite{leefer-2009}).

Consider a restricted Hilbert space of two hyperfine states, called  $1$ (``spin up'') and $2$ (``spin down''), 
of a point-like magnetic atom of mass $m$ and magnetic moment, $\v{M}$, with  
components $M_i=\mu \sigma^i$ ($i=1,2,3$) and $\vec{\sigma}$ are the usual 
spin-$\nicefrac{1}{2}$ Pauli matrices (the factor of $\hbar/2$ is absorbed in the 
definition of $\mu$). The Hamiltonian is
\beq
\hat{H} &=& \int d^3 x \psi_{\alpha}^{\dagger}(\v{x})
\left(-\frac{\hbar^2\nabla^2}{2m}\right)\psi_{\alpha}(\v{x}) + \frac{1}{2} \int d^3 x d^3 x' \nn \\
&& \times \psi_{\alpha}^{\dagger}(\v{x})\psi_{\beta}^{\dagger}(\v{x}')
U_0(\v{x},\v{x}')_{\alpha\alpha';\beta\beta'}\psi_{\beta'}(\v{x}')\psi_{\alpha'}(\v{x}) 
\eeq
where the fields $\psi_{\alpha}(\v{x})$ destroy fermions on spin state with $z$-component 
$\alpha=1,2$ at position $\v{x}$. We consider the model interaction is of the form
\beq
U_{0}(\v{x},\v{x}')_{\alpha \alpha'; \beta \beta'} &=&
\frac{\mu^2}{r^3} \sigma^{i}_{\alpha\alpha'} ( \delta_{ij} - 3 \uv{r}_i \uv{r}_j )\sigma^{j}_{\beta\beta'} \nn \\
 && \hspace{20pt}+\c g\c\delta_{\alpha\alpha'}\delta_{\beta\beta'}\delta(\v{r})  
\eeq
where $\v{r}\equiv (\v{x}-\v{x}')/|\v{x}-\v{x}'|$ and
$\uv{r}$ is a unit vector in the direction of $\v{r}$. The last (ultra-local) term represents the short-range isotropic 
(contact) interactions. It only affects the spin-triplet channel and we denote by $g$ the associated coupling constant. The Fourier transform 
of the bare two body interaction is $ (4\pi\mu^2/3)\sigma^{i}_{\alpha\alpha'}(3 
\uv{q}_{i}\uv{q}_{j} -\delta_{ij})\sigma^{j}_{\beta\beta'} + g \delta_{\alpha\alpha'}
\delta_{\beta\beta'}$. The Hamiltonian is  invariant under \textit{simultaneous} 
$SU(2)$ transformations in spin space and $SO(3)$ rotations in real space. This 
mixing of orbital and spin degrees of freedom is, in essence, what relates
the distortions of the shape of FS and the spin polarization.

\begin{figure}[hbt]
\subfigure{\includegraphics[width=0.3\textwidth]{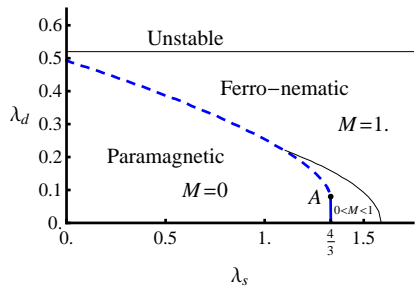}}
\caption{(color online)  Schematic phase diagram as a function of $\lambda_s = g n / \epsilon_F$ 
and $\lambda_d = n \mu^2 / \epsilon_F$. 
It has a paramagnetic phase ($M=0$), and a ferro-nematic phase with partial 
($0<M<1$) and full polarization ($M=1$).  Dashed (full) curve: first (second) order phase transitions; $A$: tricritical point.}
\label{fig:phasediagram}
\end{figure}

The ferro-nematic state breaks simultaneously the rotational invariance 
in spin space and in real space of the Hamiltonian, and its order parameters reflect this 
pattern of symmetry breaking. The order parameters are: a) the local 
magnetization vector $M_a$ ($a=x,y,z$) that measures the spin polarization, 
b) the nematic order parameter, a $3\times 3$ symmetric traceless matrix, 
$Q_{ij}$ ($i,j=x,y,z$) that measures the breaking of rotational invariance 
in space, and c) the generalized ``nematic-spin-nematic'' order parameter 
$ Q_{ij}^{ab}$, a tensor symmetric and traceless on the spatial ($i,j=x,y,z$) 
and spin ($a,b=x,y,z$) components, that measures the breaking of both 
symmetries \cite{nematic-spin-nematic}. General order parameters of the 
latter (nematic-spin-nematic) type were considered by Wu \textit{et. al.} 
\cite{wu-2007} who gave a detailed description  in 2D 
systems and partially in 3D systems. The ferro-nematic state has an 
unbroken uniaxial symmetry in real space. It is a 3D generalization 
of the quadrupolar $\alpha$ phase of Ref.\cite{wu-2007}. 

An intuitive way to describe these phases (and their order parameters) in a Fermi system is in terms 
of spontaneously deformed Fermi surfaces \cite{Oganesyan2001,
wu-2007,Fregoso-2009}. A ferromagnetic state  is isotropic in real 
space and has a spherical FS of unequal size for both 
spin polarizations. The nematic phase is isotropic in spin space, and 
its FS has a uniaxial distortion  in real space, with the up and down 
spin FS being identical in shape and size. For small values of the 
order parameter, the distorted FS are ellipsoids with an eccentricity 
determined by the magnitude of the order parameter $Q_{ij}$. Due to the mixing 
of orbital and spin degrees of freedom by the dipolar interaction, 
ferromagnetism causes the FS to distort, thus driving the system into an 
uniaxial nematic state. Since the state is ferromagnetic, the up (down) 
FS is a prolate (oblate) revolution ellipsoid. 
Both FS are collinear, and have unequal distortion and volume. Hence, 
the nematic-spin-nematic order parameter $Q_{ij}^{ab}$ has a finite value. 
We find that all three order orders are present even for arbitrary small
values of the dipolar coupling, where the ferromagnetic is the strongest,
the nematic intermediate, and the nematic-spin-nematic the weakest.

In this paper we use a Hartree-Fock (HF) variational wave function to determine 
the phase diagram as a function of the density $n$, the (dimensionless) 
dipolar coupling constant $\lambda_d$ and of the (dimensionless) local exchange coupling 
$\lambda_s=g n/\epsilon_F$. We consider an infinite system and significant finite 
size effects, such as the trap potential and the associated inhomogeneity of 
the gas, are not considered but can be included using the Thomas-Fermi approximation. 
We find two phases: a) an isotropic unpolarized state and b) a ferro-nematic phase.
As in the conventional theory of the Stoner transition, we find that the values of 
$\lambda_s$ and $\lambda_d$ on the phase boundary are of order unity. In this regime, 
a HF wave function can only yield qualitative results, such as the broad 
structure of the phase diagram, but its is not not expected to be quantitatively accurate. 
Even within mean field theory, HF may miss important physics; a recent 
study\cite{Duine2005} found that to second order in $g$ the usual 
continuous Stoner phase transition can turn first order.

We  take a variational HF wave function of the form of a 
Slater determinant describing a state in which the spin up and down 
Fermi surfaces are spontaneously deformed away from their non-interacting 
spherical shape. We will not consider other interesting states with more complex order, such as  biaxial\cite{Fregoso-2009} and its generalizations. Since we are interested in magnetism 
we allow for the volume of the up and down Fermi surfaces to change 
as well. This results in a (``Thomas-Fermi'' like) distribution function 
of fermions in momentum space with 4 variational parameters, 
$k_{F1}$, $k_{F2}$  $\alpha_1$ and $\alpha_2$. We keep the total particle 
density $n=n_1+n_2$ fixed,\cite{Miyakawa2008,Sogo}
\beq
n_{\sigma\v{k}}= \Theta\bigg(k_{F\sigma}^2 - \alpha_{\sigma}^{-1}(k^2_x + k^2_y) - \alpha_{\sigma}^2 k^2_z\bigg) 
\label{eq:distribution}
\eeq
where $\Theta(x)$ is the step function and $\sigma=1,2$. 
If $\alpha_\sigma=1$ both FS's are spheres. Eq.\eqref{eq:distribution} has the property that  $V^{-1}\sum_{\v{k}} n_{\sigma\v{k}} = k_{F\sigma}^3/(6\pi^2)\equiv n_\sigma$; the total density does not depend on
the FS distortion parameters $\alpha_{\sigma}$, with
$\alpha_\sigma>1$(oblate), $\alpha_\sigma<1$ (prolate).

Computing the energy in HF we obtain an expression for 
the ground state energy density in terms of the distribution 
functions of spin up and down particles $n_{\sigma,\v{k}}= \langle  
c^{\dagger}_{\sigma\v{k}}c_{\sigma\v{k}}\rangle$, and $\sum_\v{k} n_{\sigma,\v{k}}=N_\sigma$,
\begin{eqnarray}
\mathcal{E}_{int} &=&\c g\c n_1  n_2 + \frac{2\pi\mu^2}{3 V^2}\sum_{\v{k},\v{k}'} (1- 3 \cos^2\theta_{\v{k}-\v{k}'} )\nn \\
&&\times  \big(n_{1\v{k}'} n_{1\v{k}}  
-2 n_{1\v{k}'} n_{2\v{k}} + n_{2\v{k}'} n_{2\v{k}} \big) 
\end{eqnarray}
where $\theta_{\bf k -\bf k^\prime}$ labels the direction of $\bf k - \bf k^\prime$ 
with respect to the $z$-axis. $\mathcal{E}_{int}$ 
is a function of the magnetization $M \equiv (n_1-n_2)/n$, the density $n$,
and the ratio of the undistorted Fermi surface volumes is $r^3= k_{F1}^3/k_{F2}^3= (1+M)/(1-M)$.
The total energy \textit{density} $\mathcal{E}$ is 
\beq
\mathcal{E}_{kin} &=& \frac{C_1 \hbar^2}{3 m} \left[n_1^{5/3}\left(2\alpha_1 + \frac{1}{\alpha_1^2}\right) + n_2^{5/3}\left(2\alpha_2 + \frac{1}{\alpha_2^2}\right)\right] \nn \\
\mathcal{E}_{int} &=& g n_1 n_2 - \frac{\pi\mu^2}{3}\bigg[ n_1^2 I(\alpha_1)- 2 n_1 n_2 \c I(\alpha_1,\alpha_2,M) \nn \\
&&\hspace{65pt}+\c  n_2^2 I(\alpha_2)\bigg] 
\eeq
where $C_1 = (3/10)(6\pi^2)^{2/3}$. The integrals $I(\alpha)$ and $I(\alpha_1,\alpha_2,M)$ are given by
\begin{equation}
I(\alpha) = -2 - \frac{6}{\alpha^3 -1}- \frac{3\arccos\alpha^{3/2}}{(\alpha^{-1}- \alpha^2)^{3/2}} 
\end{equation}
\begin{widetext}
\begin{eqnarray}
I(\alpha_1,\alpha_2,M)&\equiv& \frac{9}{8\pi^2}\int d^3 k_1  d^3 k_2\c \Theta\left( 1- \alpha_1^{-1}(k_{1x}^2+k_{1y}^2) -\alpha_1^2 k_{1z}^2 \right)
\Theta\left( 1- \alpha_2^{-1}(k_{2x}^2+k_{2y}^2) -\alpha_2^2 k_{2z}^2 \right) \left(3 \cos^2\theta_{r\v{k}_1-\v{k}_2}-1\right)  \nonumber \\
&&
\end{eqnarray}
The total energy $\mathcal{E}$ becomes
\beq
\frac{\mathcal{E}}{\mathcal{E}_0} &=& 
\frac{1}{6} \left[(1 + M)^{5/3}\left(2\alpha_1 + \frac{1}{\alpha_1^2}\right) + (1 - M)^{5/3}\left(2\alpha_2 + 
\frac{1}{\alpha_2^2}\right)\right] \nn \\
&-& \frac{5\pi}{36} \lambda_d \bigg[ (1+M)^2 I(\alpha_1) - 2 (1-M^2) I(\alpha_1,\alpha_2,M) + (1-M)^2 I(\alpha_2)\bigg]  
+ \frac{5}{12}\lambda_s(1-M^2)  
\label{eqn:energy_gen}
\eeq
\end{widetext}
where $\mathcal{E}_0 = (2 C_1\hbar^2 n^{5/3})/(2^{5/3} m)$. Numerically minimizing 
the energy we obtain the phase diagram in the couplings $\lambda_s$ and $\lambda_d$ 
of  Fig. \ref{fig:phasediagram}. 
\begin{figure}[b]
\subfigure[\; Magnetization]{\includegraphics[width=0.26\textwidth]{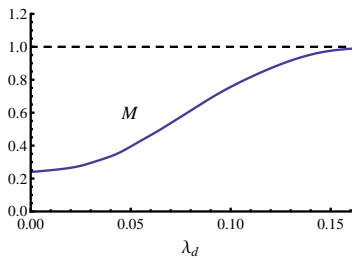}}
\subfigure[\; FS distortions]{\includegraphics[width=0.26\textwidth]{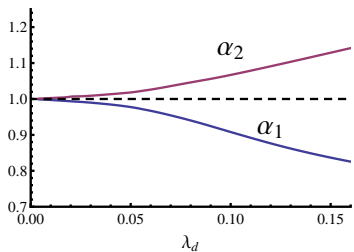}}
\caption{(color online)  Magnetization $M$ and Fermi surface 
distortion parameters $\alpha_1$ and $\alpha_2$ vs the dimensionless dipolar coupling 
$\lambda_d$ for an s-wave coupling $\lambda_s = 1.34$.}
\label{fig:a_vs_lambda}
\end{figure} 
\begin{figure}[t!]
\subfigure[$ \lambda_s=1.34$,
$\lambda_d=0.04$]{\includegraphics[width=0.23\textwidth]{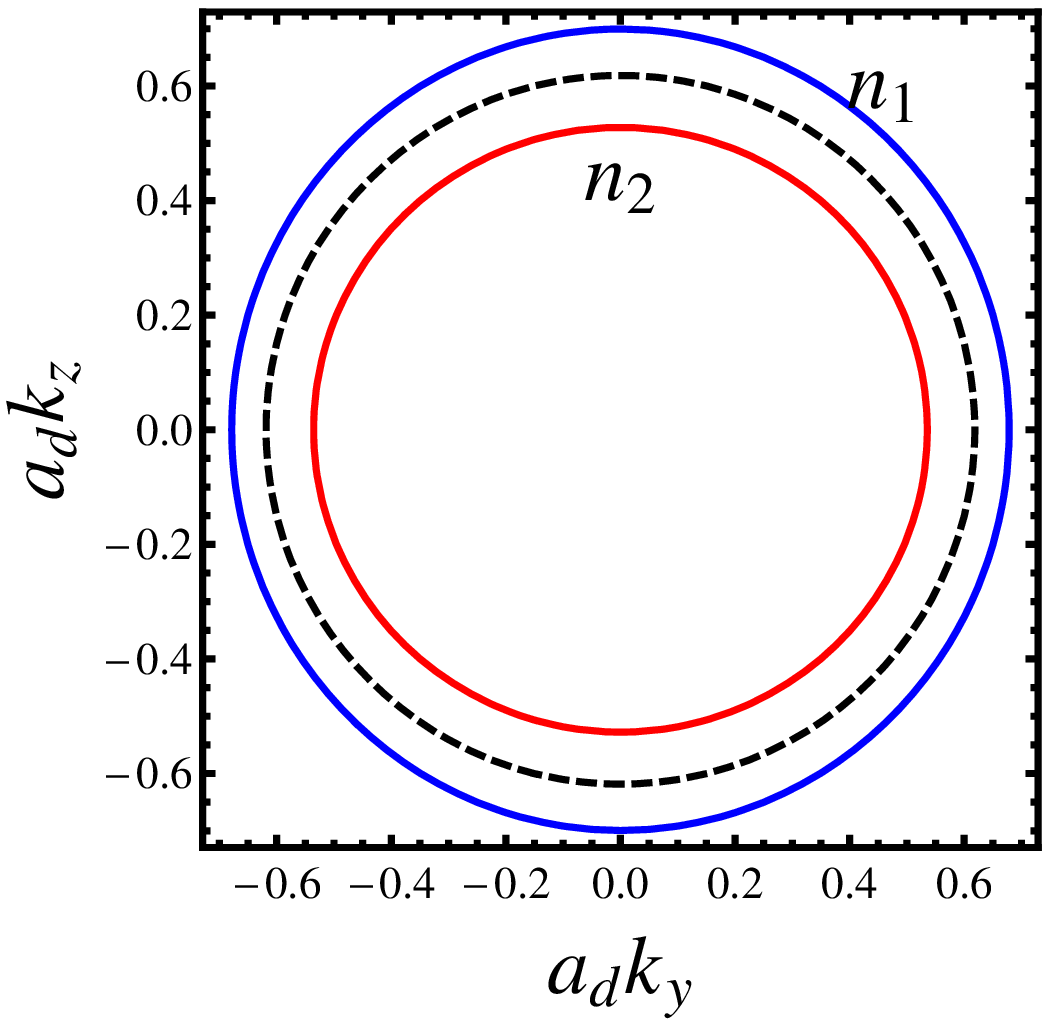}}
\subfigure[$ \lambda_s=1.34$,
$\lambda_d=0.08$]{\includegraphics[width=0.23\textwidth]{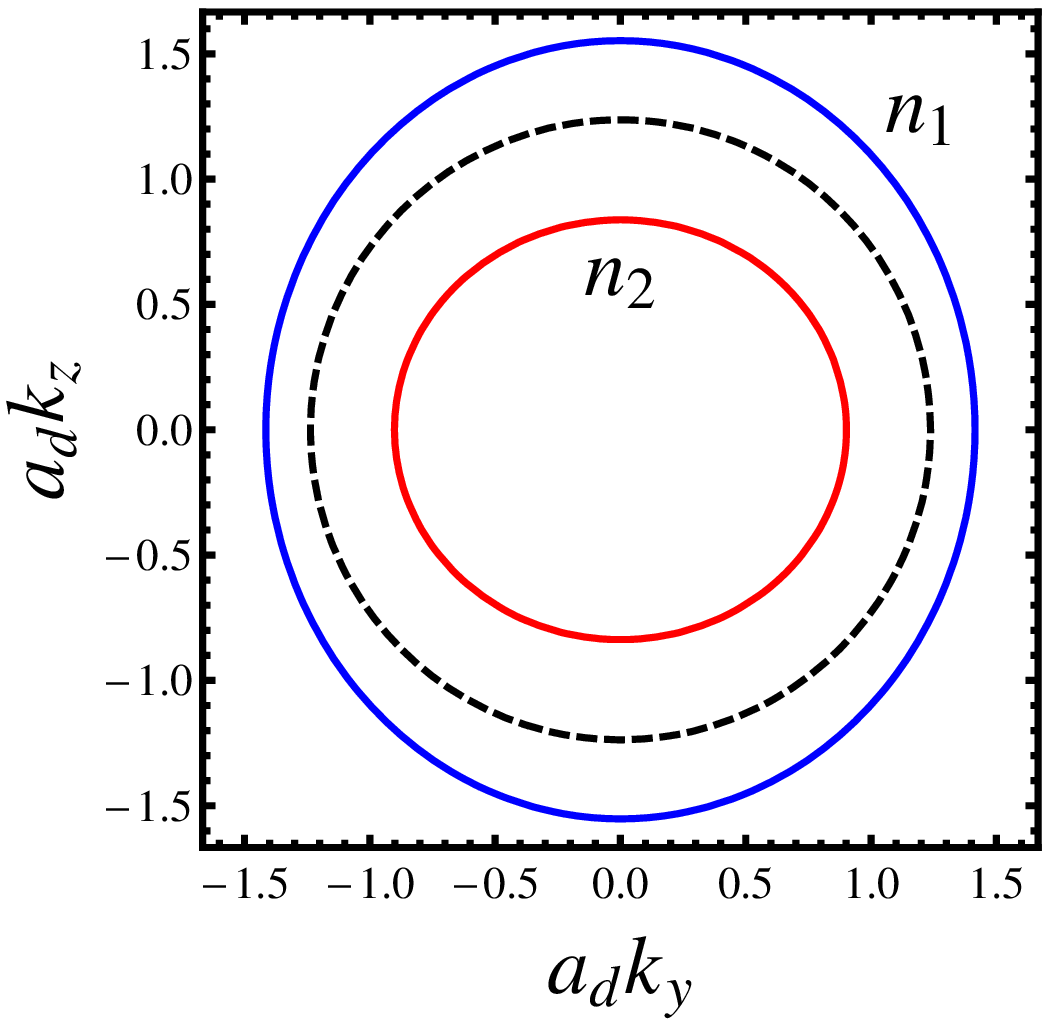}}
\subfigure[$ \lambda_s=1.37$,
$\lambda_d=0.10$]{\includegraphics[width=0.23\textwidth]{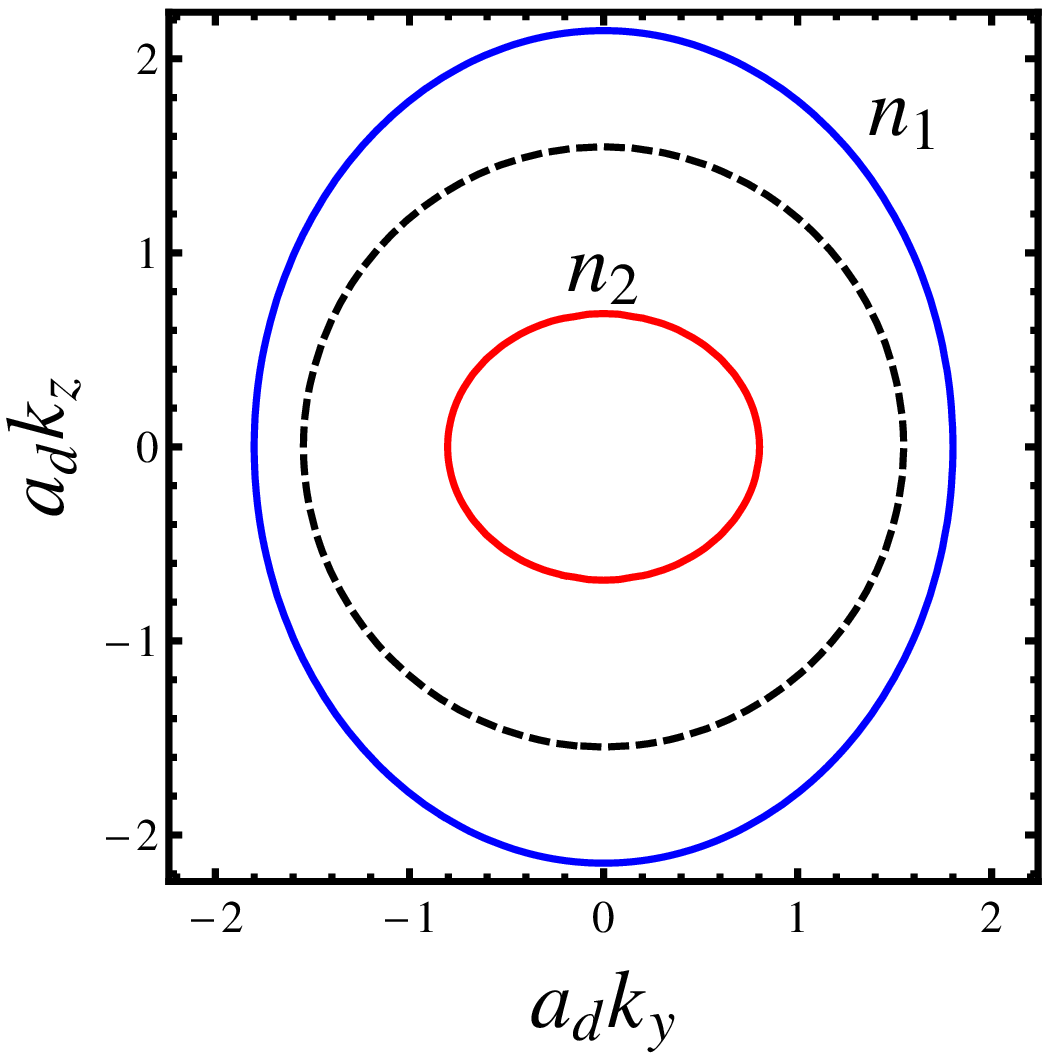}}
\subfigure[$ \lambda_s=1.37$,
$\lambda_d=0.14$]{\includegraphics[width=0.23\textwidth]{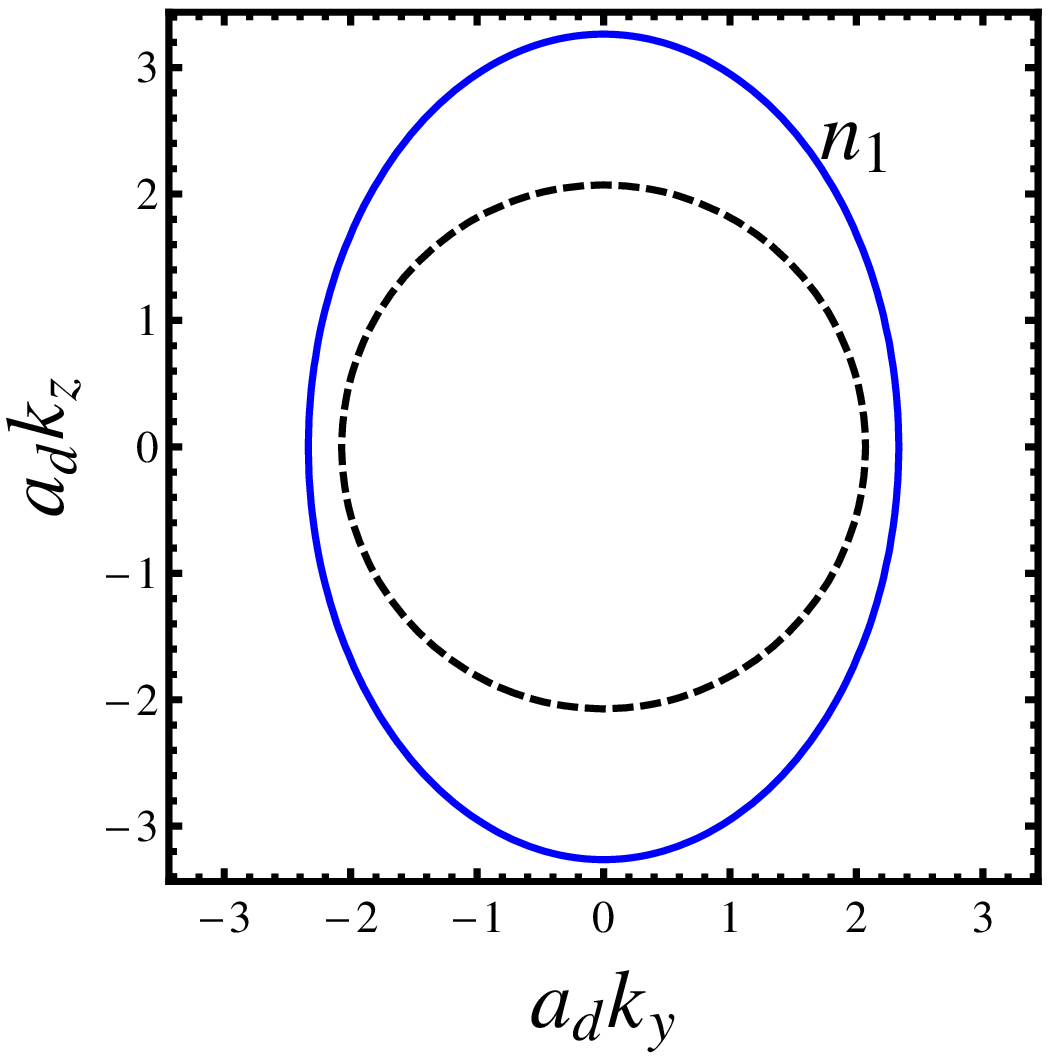}}
\caption{(color online) a-d: Spin up and down FS, labeled by $n_1$ and $n_2$, for  
several values of $\lambda_s$ and $\lambda_d$ and fixed particle density $n$.
$a_d=m\mu^2/\hbar^2$ is the dipole-dipole interaction length scale. Dashed lines: Free fermion FS.
}
\label{fig:Fermi-surfaces}
\end{figure}

The phase diagram shown in Fig. \ref{fig:phasediagram} exhibits a ferro-nematic phase and a paramagnetic phase, 
separated by a phase boundary consisting of a line of 1st order transitions that meets a line of continuous transitions at a tricritical point 
(labeled by $A$.) As expected the ferro-nematic state becomes more accessible as the $s$-wave coupling increases.  
This phase is  fully polarized ($M=1$) for most of the phase diagram, except for a small region where the polarization is partial, $0<M<1$.
In the ferro-nematic phase with partial polarization the up and down Fermi surfaces are unequally distorted, while in the fully polarized regime only
one distorted up FS exists. Partial magnetization in the conventional Stoner transition occurs for $\lambda_d=0$ and $ 4/3< \lambda_s < 2^{2/3}$, where the up and down FS become distorted even for arbitrarily small values of the dipolar coupling (see Fig. \ref{fig:a_vs_lambda}.)  
From the structure of the free energy for small values of 
$Q_1=\alpha_1-1$, $Q_2=\alpha_2-1$ and $M$ (valid in the vicinity of the continuous transition) we see that the dipolar interaction leads to a 
leading term of the form $M(Q_1-Q_2)$, which is invariant under $1\leftrightarrow 2$.  
Since the ferromagnetic state is already favored by 
the contact term, this term also favors $Q_1<0$ and $Q_2>0$.

In the ferro-nematic phase the $SU(2)$ spin symmetry of the Hamiltonian is broken down to the residual 
$U(1) \cong SO(2)$ invariance of this uniaxial state. The equilibrium FS of the up 
and down spin components are shown in Fig. \ref{fig:Fermi-surfaces} for several values of of
the coupling constants. Both FS's are invariant under $SO(2)$ rigid rotations about 
$\v{M}$.

\begin{figure}[t!]
\subfigure{\includegraphics[width=0.35\textwidth]{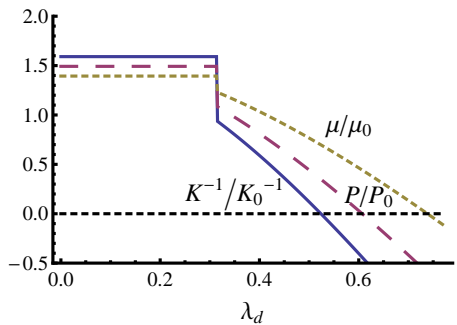}}
\caption{(color online) Pressure (broken), bulk compressibility (full) and chemical potential (dashed) 
(normalized to their values at the non-interacting Fermi gas) vs the 
dimensionless dipolar coupling $\lambda_d$, for $\lambda_s = 0.8$. At $\lambda_d \sim 0.3$ there is a 1st order paramagnetic-ferro-nematic transition. The gas becomes unstable at $\lambda_d \simeq 0.52$.}
\label{fig:thermo}
\end{figure}

The total energy, the equilibrium values of the FS distortions and the magnetization 
are functions of the particle density $n=N/V$. The pressure $P=-(\partial E/\partial V)_N$, 
chemical potential $\mu = (\partial E/\partial N)_V$ and the compressibility 
$K^{-1}= n(\partial P/\partial n)$ can be computed straightforwardly, and  decrease monotonically as the dipolar coupling 
increases, see Fig. \ref{fig:thermo}. The compressibility vanishes at  $\lambda_d^c\simeq 0.52$
\cite{Sogo,comment_sogo} where the Fermi gas becomes (formally) unstable to collapse\cite{instability}. 
%

In current experiments on cold atoms it is possible to prepare a two-component 
Fermi gas out of the hyperfine manifold of the atom. One can imagine, 
a two-component dipolar Fermi gas made out of, say, the hyperfine manifold of strongly 
magnetic atom such as Dy. For $^{163}$Dy with a density of $10^{13}$cm$^{-3}$, we estimate 
$\lambda_{d}\simeq 0.01$ \cite{caveat,chromium}. 
The recent experimental observation of itinerant (Stoner) ferromagnetism in ultra-cold gases of $^6$Li atoms\cite{ketterle-2009}
opens the possibility to detect the ferro-nematic state in the laboratory possibly by tuning the $s$-wave scattering amplitude.
Depending on the strength of the dipole moment and of the $s$-wave coupling, the FS for the two components may 
differ considerably from the spherical shape, and free expansion experiments 
may be able to provide signatures of this state. Ferromagnetism in cold dilute 
systems interacting only by s-wave contact pseudo-potential has been difficult to observe as the spin states of 
fermions are conserved separately on the time scales of the experiments\cite{Duine2005}. 
However, a ferro-nematic state can be set up experimentally in systems with a population imbalance of hyperfine states, which
may also exhibit other  anisotropic and inhomogeneous phases. In actual experiments the trap potential is anisotropic and weakly inhomogeneous. We estimate that such effects do not affect our main results provided the trap potential aspect ratios depart from 1 by perhaps up to about 30\%. Nevertheless,  the anisotropy  acts as a weak symmetry breaking field, that orients the ferro-nematic order. 

In this work we have shown the existence a new phase of matter, the
ferro-nematic Fermi fluid, a ground state of a dipolar Fermi gas with short range interactions
with a spontaneous magnetization and long range orientational order. In this state, the up 
and down FS manifolds have unequal shapes and volumes. Since rotational invariance in real space 
and in spin space is simultaneously spontaneously broken in this state, it supports a rich 
spectrum of Goldstone excitations. As a result the fluid is an optically anisotropic medium 
whose properties that may be detected by light scattering experiments.

We thank B. L. Lev, B. Hsu, K. Sun, and B. Uchoa for many discussions. EF thanks the Kavli Institute 
for Theoretical Physics (KITP) of the University of California Santa Barbara for hospitality. 
This work was 
supported in part by the National Science Foundation under the grants DMR 0758462 (EF) at 
the University of Illinois, and  PHY05-51164 (EF) at  KITP, and by the Office of Science, 
U.S. Department of Energy under contracts DE-FG02-91ER45439 (EF, BMF) through the University 
of Illinois Frederick Seitz Materials Research Laboratory.


\begin{thebibliography}{23}
\expandafter\ifx\csname natexlab\endcsname\relax\def\natexlab#1{#1}\fi
\expandafter\ifx\csname bibnamefont\endcsname\relax
  \def\bibnamefont#1{#1}\fi
\expandafter\ifx\csname bibfnamefont\endcsname\relax
  \def\bibfnamefont#1{#1}\fi
\expandafter\ifx\csname citenamefont\endcsname\relax
  \def\citenamefont#1{#1}\fi
\expandafter\ifx\csname url\endcsname\relax
  \def\url#1{\texttt{#1}}\fi
\expandafter\ifx\csname urlprefix\endcsname\relax\def\urlprefix{URL }\fi
\providecommand{\bibinfo}[2]{#2}
\providecommand{\eprint}[2][]{\url{#2}}

\bibitem[{\citenamefont{Miyakawa et~al.}(2008)\citenamefont{Miyakawa, Sogo, and
  Pu}}]{Miyakawa2008}
\bibinfo{author}{\bibfnamefont{T.}~\bibnamefont{Miyakawa}},
  \bibinfo{author}{\bibfnamefont{T.}~\bibnamefont{Sogo}}, \bibnamefont{and}
  \bibinfo{author}{\bibfnamefont{H.}~\bibnamefont{Pu}}, \bibinfo{journal}{Phys.
  Rev. A} \textbf{\bibinfo{volume}{77}}, \bibinfo{pages}{061603(R)}
  (\bibinfo{year}{2008}).

\bibitem[{\citenamefont{Sogo et~al.}(2009)\citenamefont{Sogo, He, Miyakawa, Yi,
  Lu, and Pu}}]{Sogo}
\bibinfo{author}{\bibfnamefont{T.}~\bibnamefont{Sogo}},
  \bibinfo{author}{\bibfnamefont{L.}~\bibnamefont{He}},
  \bibinfo{author}{\bibfnamefont{T.}~\bibnamefont{Miyakawa}},
  \bibinfo{author}{\bibfnamefont{S.}~\bibnamefont{Yi}},
  \bibinfo{author}{\bibfnamefont{H.}~\bibnamefont{Lu}}, \bibnamefont{and}
  \bibinfo{author}{\bibfnamefont{H.}~\bibnamefont{Pu}}, \bibinfo{journal}{New
  J. Phys.} \textbf{\bibinfo{volume}{11}}, \bibinfo{pages}{055017}
  (\bibinfo{year}{2009}).

\bibitem[{\citenamefont{Fregoso et~al.}(2009)\citenamefont{Fregoso, Sun,
  Fradkin, and Lev}}]{Fregoso-2009}
\bibinfo{author}{\bibfnamefont{B.~M.} \bibnamefont{Fregoso}},
  \bibinfo{author}{\bibfnamefont{K.}~\bibnamefont{Sun}},
  \bibinfo{author}{\bibfnamefont{E.}~\bibnamefont{Fradkin}}, \bibnamefont{and}
  \bibinfo{author}{\bibfnamefont{B.~L.} \bibnamefont{Lev}},
  \bibinfo{journal}{New J. Phys.} \textbf{\bibinfo{volume}{11}},
  \bibinfo{pages}{103003} (\bibinfo{year}{2009}).

\bibitem[{\citenamefont{Kivelson et~al.}(1998)\citenamefont{Kivelson, Fradkin,
  and Emery}}]{Kivelson1998}
\bibinfo{author}{\bibfnamefont{S.~A.} \bibnamefont{Kivelson}},
  \bibinfo{author}{\bibfnamefont{E.}~\bibnamefont{Fradkin}}, \bibnamefont{and}
  \bibinfo{author}{\bibfnamefont{V.~J.} \bibnamefont{Emery}},
  \bibinfo{journal}{Nature} \textbf{\bibinfo{volume}{393}},
  \bibinfo{pages}{550} (\bibinfo{year}{1998}).

\bibitem[{\citenamefont{Oganesyan et~al.}(2001)\citenamefont{Oganesyan,
  Kivelson, and Fradkin}}]{Oganesyan2001}
\bibinfo{author}{\bibfnamefont{V.}~\bibnamefont{Oganesyan}},
  \bibinfo{author}{\bibfnamefont{S.~A.} \bibnamefont{Kivelson}},
  \bibnamefont{and} \bibinfo{author}{\bibfnamefont{E.}~\bibnamefont{Fradkin}},
  \bibinfo{journal}{Phys. Rev. B} \textbf{\bibinfo{volume}{64}},
  \bibinfo{pages}{195109} (\bibinfo{year}{2001}).

\bibitem[{\citenamefont{Sun et~al.}(2008)\citenamefont{Sun, Fregoso, Lawler,
  and Fradkin}}]{Sun2008}
\bibinfo{author}{\bibfnamefont{K.}~\bibnamefont{Sun}},
  \bibinfo{author}{\bibfnamefont{B.~M.} \bibnamefont{Fregoso}},
  \bibinfo{author}{\bibfnamefont{M.~J.} \bibnamefont{Lawler}},
  \bibnamefont{and} \bibinfo{author}{\bibfnamefont{E.}~\bibnamefont{Fradkin}},
  \bibinfo{journal}{Phys. Rev. B} \textbf{\bibinfo{volume}{78}},
  \bibinfo{pages}{085124} (\bibinfo{year}{2008}).

\bibitem[{\citenamefont{Doniach and Sondheimer}(1998)}]{Doniach-1974}
\bibinfo{author}{\bibfnamefont{S.}~\bibnamefont{Doniach}} \bibnamefont{and}
  \bibinfo{author}{\bibfnamefont{E.~H.} \bibnamefont{Sondheimer}},
  \emph{\bibinfo{title}{Green's Functions For Solid State Physicists}}
  (\bibinfo{publisher}{Imperial College Press}, \bibinfo{year}{1998}).

\bibitem[{mah()}]{mahanti-comment}
\bibinfo{note}{Ref.\cite{Mahanti2007} used a similar variational state to argue
  that the dipolar Fermi gas is fully polarized.}
  
  \bibitem[{\citenamefont{Mahanti and Jha}(2007)}]{Mahanti2007}
\bibinfo{author}{\bibfnamefont{S.~D.} \bibnamefont{Mahanti}} \bibnamefont{and}
  \bibinfo{author}{\bibfnamefont{S.~S.} \bibnamefont{Jha}},
  \bibinfo{journal}{Phys. Rev. E} \textbf{\bibinfo{volume}{76}},
  \bibinfo{pages}{062101} (\bibinfo{year}{2007}).


\bibitem[{\citenamefont{Brochard and de~Gennes}(1970)}]{Brochard1970}
\bibinfo{author}{\bibfnamefont{F.}~\bibnamefont{Brochard}} \bibnamefont{and}
  \bibinfo{author}{\bibfnamefont{P.~G.} \bibnamefont{de~Gennes}},
  \bibinfo{journal}{J. Physique (Paris)} \textbf{\bibinfo{volume}{31}},
  \bibinfo{pages}{691} (\bibinfo{year}{1970}).

\bibitem[{\citenamefont{Seul and Andelman}(1995)}]{seul-1995}
\bibinfo{author}{\bibfnamefont{M.}~\bibnamefont{Seul}} \bibnamefont{and}
  \bibinfo{author}{\bibfnamefont{D.}~\bibnamefont{Andelman}},
  \bibinfo{journal}{Science} \textbf{\bibinfo{volume}{267}},
  \bibinfo{pages}{477} (\bibinfo{year}{1995}).

\bibitem[{\citenamefont{Fomin et~al.}(1978)\citenamefont{Fomin, Pethick, and
  Serene}}]{Fomin1978}
\bibinfo{author}{\bibfnamefont{I.~A.} \bibnamefont{Fomin}},
  \bibinfo{author}{\bibfnamefont{C.~J.} \bibnamefont{Pethick}},
  \bibnamefont{and} \bibinfo{author}{\bibfnamefont{J.~W.}
  \bibnamefont{Serene}}, \bibinfo{journal}{Phys Rev. Lett.}
  \textbf{\bibinfo{volume}{40}}, \bibinfo{pages}{1144} (\bibinfo{year}{1978}).

\bibitem[{\citenamefont{Leefer et~al.}(2009)\citenamefont{Leefer, Cingoz, and
  Budker}}]{leefer-2009}
\bibinfo{author}{\bibfnamefont{N.}~\bibnamefont{Leefer}},
  \bibinfo{author}{\bibfnamefont{A.}~\bibnamefont{Cingoz}}, \bibnamefont{and}
  \bibinfo{author}{\bibfnamefont{D.}~\bibnamefont{Budker}},
  \bibinfo{journal}{Optics Lett.} \textbf{\bibinfo{volume}{34}},
  \bibinfo{pages}{2548} (\bibinfo{year}{2009}).

\bibitem[{nem()}]{nematic-spin-nematic}
\bibinfo{note}{{A nematic-spin-nematic \cite{Kivelson2003,wu-2007} is a 2D
  state invariant under a rotation by $\pi$ followed by a spin flip. The
  uniaxial 3D ferro-nematic state does not have this property, as its up FS and
  down FS are not equivalent under a rotation.}}
  
  \bibitem[{\citenamefont{Kivelson et~al.}(2003)\citenamefont{Kivelson, Bindloss,
  Fradkin, Oganesyan, Tranquada, Kapitulnik, and Howald}}]{Kivelson2003}
\bibinfo{author}{\bibfnamefont{S.~A.} \bibnamefont{Kivelson}},
  \bibinfo{author}{\bibfnamefont{I.~P.} \bibnamefont{Bindloss}},
  \bibinfo{author}{\bibfnamefont{E.}~\bibnamefont{Fradkin}},
  \bibinfo{author}{\bibfnamefont{V.}~\bibnamefont{Oganesyan}},
  \bibinfo{author}{\bibfnamefont{J.~M.} \bibnamefont{Tranquada}},
  \bibinfo{author}{\bibfnamefont{A.}~\bibnamefont{Kapitulnik}},
  \bibnamefont{and} \bibinfo{author}{\bibfnamefont{C.}~\bibnamefont{Howald}},
  \bibinfo{journal}{Rev. Mod. Phys.} \textbf{\bibinfo{volume}{75}},
  \bibinfo{pages}{1201} (\bibinfo{year}{2003}).



\bibitem[{\citenamefont{Wu et~al.}(2007)\citenamefont{Wu, Sun, Fradkin, and
  Zhang}}]{wu-2007}
\bibinfo{author}{\bibfnamefont{C.}~\bibnamefont{Wu}},
  \bibinfo{author}{\bibfnamefont{K.}~\bibnamefont{Sun}},
  \bibinfo{author}{\bibfnamefont{E.}~\bibnamefont{Fradkin}}, \bibnamefont{and}
  \bibinfo{author}{\bibfnamefont{S.~C.} \bibnamefont{Zhang}},
  \bibinfo{journal}{Phys. Rev. B} \textbf{\bibinfo{volume}{75}},
  \bibinfo{pages}{115103} (\bibinfo{year}{2007}).

\bibitem[{\citenamefont{Duine and MacDonald}(2005)}]{Duine2005}
\bibinfo{author}{\bibfnamefont{R.~A.} \bibnamefont{Duine}} \bibnamefont{and}
  \bibinfo{author}{\bibfnamefont{A.~H.} \bibnamefont{MacDonald}},
  \bibinfo{journal}{Phys Rev Lett} \textbf{\bibinfo{volume}{95}},
  \bibinfo{pages}{230403} (\bibinfo{year}{2005}).

\bibitem[{com()}]{comment_sogo}
\bibinfo{note}{{Sogo et al \cite{Sogo} estimated that the homogeneous fully
  polarized dipolar Fermi gas becomes unstable at $C_{dd}\equiv (3\pi^2)^{2/3}
  (\lambda_d/2) \simeq 3.2$. The correct value is $C_{dd}\simeq 2.5$.}}

\bibitem[{ins()}]{instability}
\bibinfo{note}{{Short-range interactions tend to suppress this instability.}}

\bibitem[{cav()}]{caveat}
\bibinfo{note}{{Dy has $I=5/2$, $J=8$ and $11/2 \leq F \leq
  21/2$.\cite{leefer-2009} It is not described by this simple two-state
  model.}}

\bibitem[{chr()}]{chromium}
\bibinfo{note}{{Dipolar atoms such as $^{52}$Cr have been cooled and a
  Bose-Einstein condensate has been observed\cite{french}. $^{53}$Cr is a
  fermion and, if cooled to low enough temperatures, could also be a candidate
  for a ferro-nematic state.}}
  
  \bibitem[{\citenamefont{Beaufulis et~al.}(2008)\citenamefont{Beaufils,
  Chicireanu, Zanon, Laburthe-Tolra, {Mar\'echal}, Vernac, Keller, and
  Gorceix}}]{french}
\bibinfo{author}{\bibfnamefont{Q.}~\bibnamefont{Beaufils}},
  \bibinfo{author}{\bibfnamefont{R.}~\bibnamefont{Chicireanu}},
  \bibinfo{author}{\bibfnamefont{T.}~\bibnamefont{Zanon}},
  \bibinfo{author}{\bibfnamefont{B.}~\bibnamefont{Laburthe-Tolra}},
  \bibinfo{author}{\bibfnamefont{E.}~\bibnamefont{{Mar\'echal}}},
  \bibinfo{author}{\bibfnamefont{L.}~\bibnamefont{Vernac}},
  \bibinfo{author}{\bibfnamefont{J.~C.} \bibnamefont{Keller}},
  \bibnamefont{and} \bibinfo{author}{\bibfnamefont{O.}~\bibnamefont{Gorceix}},
  \bibinfo{journal}{Phys Rev. A} \textbf{\bibinfo{volume}{77}},
  \bibinfo{pages}{061601(R)} (\bibinfo{year}{2008}).


\bibitem[{\citenamefont{Jo et~al.}(2009)\citenamefont{Jo, Lee, Choi,
  Christiansen, Kim, Thywissen, Pritchard, and Ketterle}}]{ketterle-2009}
\bibinfo{author}{\bibfnamefont{G.~B.} \bibnamefont{Jo}},
  \bibinfo{author}{\bibfnamefont{Y.~R.} \bibnamefont{Lee}},
  \bibinfo{author}{\bibfnamefont{J.~H.} \bibnamefont{Choi}},
  \bibinfo{author}{\bibfnamefont{C.~A.} \bibnamefont{Christiansen}},
  \bibinfo{author}{\bibfnamefont{T.~H.} \bibnamefont{Kim}},
  \bibinfo{author}{\bibfnamefont{J.~H.} \bibnamefont{Thywissen}},
  \bibinfo{author}{\bibfnamefont{D.~E.} \bibnamefont{Pritchard}},
  \bibnamefont{and} \bibinfo{author}{\bibfnamefont{W.}~\bibnamefont{Ketterle}},
  \bibinfo{journal}{Science} \textbf{\bibinfo{volume}{325}},
  \bibinfo{pages}{1521} (\bibinfo{year}{2009}).



\end{thebibliography}

\end{document}